# Piezoelectric properties of graphene oxide from the first-principles calculations

Zhenyue Chang[1], Wenyi Yan[1], Jin Shang[2] and Jefferson Zhe Liu[1*]

[1] *Department of Mechanical and Aerospace Engineering, Monash University, Clayton, VIC 3800, Australia*

[2] *Department of Chemical and Biomolecular Engineering, The University of Melbourne, Victoria 3010, Australia.*

**Abstract**

Some highly ordered compounds of graphene oxide (GO), *e.g.*, the so-called clamped and unzipped GO, are shown to have piezoelectric responses via first-principles density functional calculations. By applying an electric field perpendicular to the GO basal plane, the largest value of in-plane strain and strain piezoelectric coefficient, $d_{31}$ are found to be 0.12% and 0.24 pm/V, respectively, which are comparable with those of some advanced piezoelectric materials. An in-depth molecular structural analysis reveals that deformation of the oxygen doping regions in the clamped GO dominates its overall strain output, whereas deformation of the regions without oxygen dopant in the unzipped GO determines its overall piezoelectric strain. This understanding explains the observed dependence of $d_{31}$ on oxygen doping rate, *i.e.*, higher oxygen concentration giving rise to a larger $d_{31}$ in the clamped GO whereas leading to a reduced $d_{31}$ in the unzipped GO. As the thinnest two-dimensional piezoelectric materials, GO has a great potential for a wide range of MEMS/NEMS actuators and sensors.

[*] Email: zhe.liu@monash.edu

There has been an immense effort towards the development of advanced actuation materials in the past decade. Actuators have been adopted in a diverse range of micro/nanoelectromechanical systems (MEMS/NEMS), including medical devices,[1] microrobotic,[2] artificial muscle,[3,4] and many other smart structures.[5,6] The widely used actuation materials nowadays have advantages and meanwhile suffer some limitations. For instance, polymers-based actuators have excellent strain output (up to ~100%) and a lightweight, but the response speed is low.[7,8] Piezoelectric actuators are capable of generating a linear strain output, however the small strain output (0.1-0.2%) and the requirement of high operating voltage restricts its use in MEMS/NEMS devices.[9,10] Recently, carbon-based materials, such as carbon nanotubes and graphene, have attracted a lot of interest in designing high performance actuators because of their unique atomic structure and excellent physical properties.[11-17] While some high performance carbon-based actuators have been shown in experiments,[18-20] a comprehensive understanding of the actuation mechanisms is vital in order to realize the full potential of these new materials.

Recently, Ong et al. used density functional theory (DFT) calculations to study the piezoelectric properties of graphene-based materials.[21] It is well known that the piezoelectric effect only exists in crystalline materials with no inversion symmetry. To break the inversion symmetry of pristine graphene, Ong et al. introduced physisorbed adatoms (e.g., Li, K, H and F) onto the graphene surface. As a result, a maximum piezoelectric linear strain around 0.15% was generated.[21,22] However, the main drawback of physisorption is the weak interactions between the adsorbed adatoms and graphene substrate. It may lead to desorption at a relative high operation temperature or under a high actuation frequency, causing the potential failure of materials and devices.[23]

Graphene oxide (GO), usually as the pre-product of synthesizing graphene,[24,25] has generated huge interest for different types of applications. The vast diversity of GO atomic structures gives rise to different electronic and mechanical properties that are potentially useful for actuation material design.[26-29] For example, recent experiments have shown highly ordered doping of oxygen (O) atoms on the hexagonal lattice of pristine graphene. Approximately 50% of the GO surfaces characterized using scanning tunneling microscope (STM) were found to comprise these novel periodic structures.[16] There are two possible O atom doping configurations: so-called clamped and unzipped (Fig. 1).[30] It is foreseeable that the doping of O atoms onto the surface with formation of strong chemical bonds will break the inversion symmetry of pristine graphene, therefore inducing piezoelectricity in GO. In this paper, we use first-principles density functional theory (DFT) to investigate piezoelectric response of GO with different structural configurations. The obtained strain piezoelectric coefficients, $d_{31}$ are compared with results of the graphene with physisorbed atoms. We also conduct an analysis to understand the structural origin of the piezoelectric strain in GO, which can explain the different dependence of $d_{31}$ on oxygen doping rate for the clamped and unzipped GO.

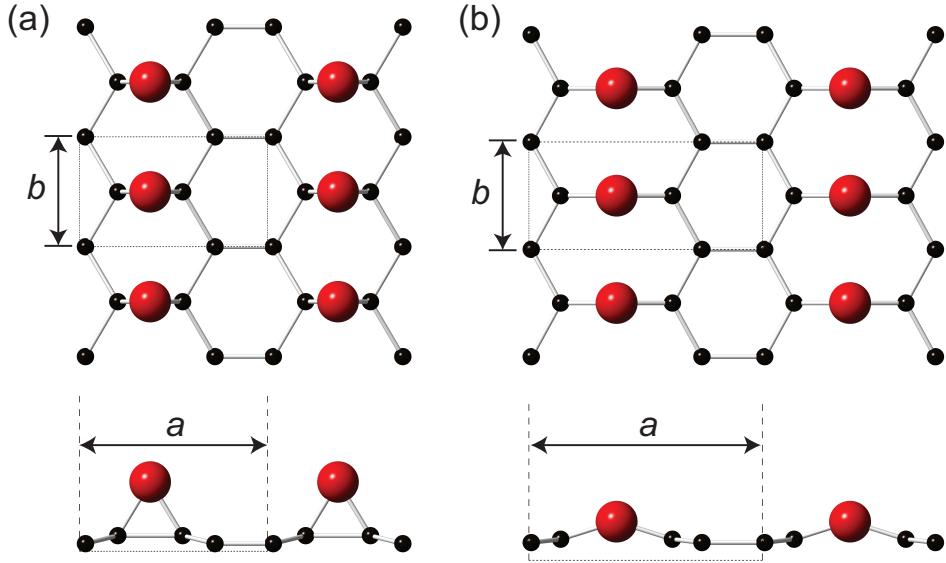

FIG. 1. Symmetrically clamped (a) and unzipped (b) GO configurations with a C/O ratio, $R_{C/O}$ of 4:1. Unit cells are depicted by dotted lines with in-plane lattice parameters shown as $a$ and $b$. The C and O atoms are represented by small black and large red spheres, respectively.

Figure 1 shows the unit cells of clamped and unzipped GO with a C/O ratio, $R_{C/O}$ of 4:1. In both types of GO molecules, the oxygen atoms form covalent bonds with the two underneath carbon atoms. For the unzipped GO, the C-C bond below oxygen atom is broken. Following the terminology from Xue *et al.*, we term them as symmetrically clamped GO (sym-clamped) and symmetrically unzipped GO (sym-unzipped), respectively.[30] With the epoxy groups attached to one side, the crystal symmetry will be transformed from a point group of *6/mmm* for pristine graphene to a non-inversion symmetric point group of *mm2*. In this study, the clamped GO with a $R_{C/O}$ of 2 or 4 was examined. Our DFT calculations showed that with $R_{C/O} > 4$, the clamped GO were unstable. For the unzipped GO, a $R_{C/O}$ of 4, 8, or 16 was studied. Noting that an unzipped GO molecule with $R_{C/O} < 4$ was unstable.

The Vienna ab initio simulation package (VASP v.5.3.3) was used to perform density functional calculations on piezoelectric responses of symmetric GO. Projector augments wave (PAW) pseudopotentials and the generalized gradient approximation (GGA) were used,[31,32] with a plane-wave cutoff energy of 800 eV. A Monkhorst-Pack gamma-centered $k$-points grid of dimensions 24×42×1 was adopted for the $C_2O$-sym-clamped and $C_4O$-sym-clamped cells, with a 10×40×1 was used for the $C_8O$-sym-unzipped GO cell. As periodic boundary conditions are employed in VASP, very thick vacuum layers were included adjacent to the GO in order to minimize interlayer interactions. An interlayer spacing of 20 Å was used throughout, which provides a good balance between computational accuracy and efforts.[33] To hold this interlayer space constant, the VASP source code was modified to allow the simulation cells to completely relax within the plane of GO, not perpendicular to the plane. In all cases, the C and O atoms were allowed to relax freely in all directions. Prior to being subject to an external electric field, all structures were fully relaxed to determine their equilibrium lattice constants. The relative change of in-plane lattice constants under an applied electric field with respect to the equilibrium values are defined as piezoelectric strains.

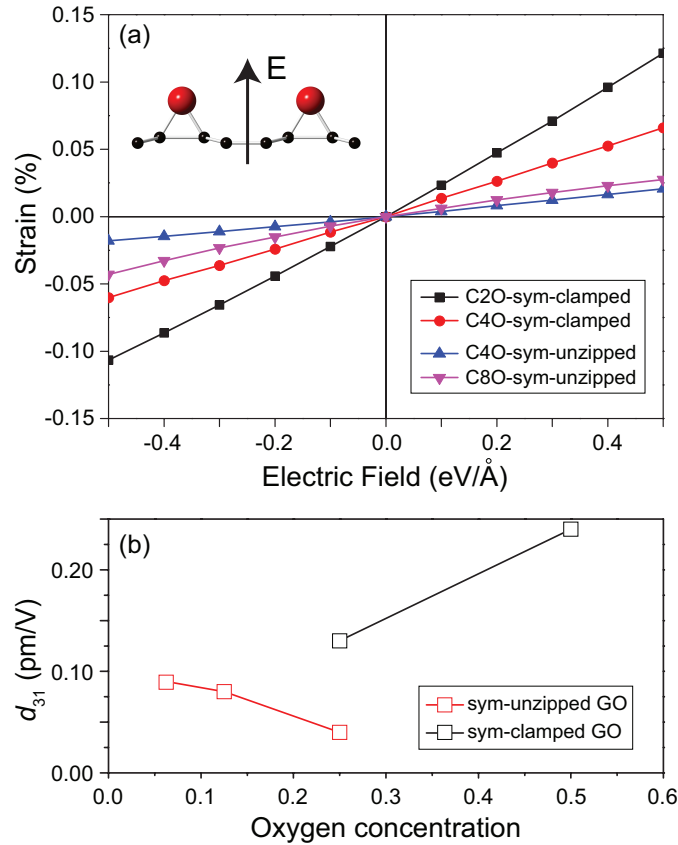

FIG. 2. (a) Piezoelectric strain of symmetric GO configurations being subject to an applied electric field perpendicular to the basal plane (inset). (b) The strain piezoelectric coefficient, $d_{31}$ as a function of oxygen doping level for sym-clamped GO and sym-unzipped GO.

Figure 2(a) shows the in-plane piezoelectric strains as a function of electric field strength from -0.5 to 0.5 eV/Å. A linear relation is observed for the sym-clamped GO molecules with $R_{C/O}$ of 2 and 4 and the sym-unzipped GO molecules with $R_{C/O}$ of 4 and 8. Note that the magnitudes of the applied electric fields are experimentally achievable in graphene-based devices.[34] Overall, the sym-clamped GO has a better strain output than the sym-unzipped GO. The maximum strain output of 0.12% at an applied filed of 0.5 eV/Å is comparable with the results from Ong *et al*. for graphene with some physisorbed adatoms (0.15%).[21]

Figure 2(b) shows that the strain piezoelectric coefficient, $d_{31}$ (*i.e.*, the slope of strain vs. electric field curve) as a function of oxygen concentration. Overall the $d_{31}$ coefficients for the clamped GO are significantly larger than those of the unzipped GO. A higher oxygen concentration in sym-clamped GO leads to a larger $d_{31}$, whereas an opposite trend is observed for the sym-unzipped GO. The maximum $d_{31}$ is obtained for $C_2O$-sym-clamped GO. Which is comparable with the maximum $d_{31}$ coefficient of 0.3 pm/V for the engineered piezoelectric graphene,[21] and the $d_{31}$ coefficients of the three dimensional piezoelectric materials like wurtzite boron nitride[35] and wurtzite GaN[36], 0.33 pm/V and 0.96 pm/V, respectively. But the $d_{31}$ value is far smaller than the most widely used piezoelectric ceramics PZT, 119 pm/V.[37] It should be noted that for piezoelectric ceramic thin films with a thickness lower than 10nm, the depolarization field generated by the accumulated surface charges will completely suppress the piezoelectric effects. Our proposed piezoelectric GO will not suffer this problem, rendering it great potentials in NEMS applications.

To gain an in-depth understanding of piezoelectric properties of sym-clamped and sym-unzipped GO, we decompose the in-plane deformation of GO molecules into two contributions. As shown in the insets of Fig. 3, the in-plane projection of interatomic distance of the two carbon atoms bonded by the oxygen atoms is defined as segment-1. The segment-2 is the in-plane projection of rest part. The total deformation, $\Delta_{tot}$ as shown in Fig. 3 is the summation of the length change in segment-1 and segment-2. For the sym-clamped cases (Fig. 3(a) and 3(b)), it is evident that the length change of segment-1, $\Delta_1$ dominates the overall piezoelectric strain output. It appears that $\Delta_1$ for $C_2O$-sym-clamped is almost twice of that of C4O-sym-clamped GO, which is consistent with increase of oxygen concentration or the increase of portions of segment-1. The length change in segment-2, $\Delta_2$ is approxiamtely the same for the two clamped GO molecules. Thus it is reasonable to understand that the strain output as well as strain piezoelectric coefficient $d_{31}$ is nearly doubled in $C_2O$-sym-clamped GO when comparing with the results of $C_4O$-sym-clamped GO (Fig. 2). In contrast, for the unzipped GO (Fig. 3(c) and 3(d)), $\Delta_1$ is virtually negligible and the deformation mainly comes from segment-2. Clearly the increase of oxygen concentration will reduce the number of segment-2 in the unzipped GO. Therefore, it is rational to understand variation trend of $d_{31}$ with respect to oxygen contration in Fig. 2(b).

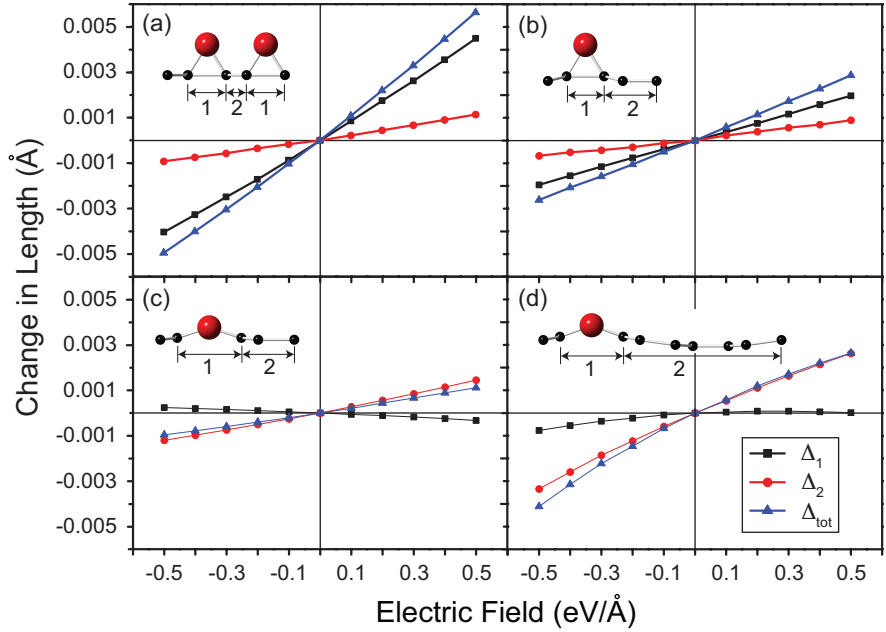

FIG. 3. Stran analysis (a) $C_2O$-sym-clamped (b) $C_4O$-sym-clamped (c) $C_4O$-sym-unzipped (d) $C_8O$-sym-unzipped. The change in length of unit cell (total), segment-1 and segment-2 are represented by blue, black and red lines respectively. Inset: side view of (a) $C_2O$-sym-clamped (b) $C_4O$-sym-clamped and (c) $C_4O$-sym-unzipped (d) $C_8O$-sym-unzipped with indication of segment-1 and 2. The C and O atoms are represented by small black and large red spheres, respectively.

With the oxygen concentration of GO reduced to 0.0625 (*i.e.*, $R_{C/O}$ = 16), some distinctive electromechanical properties were observed. Figure 4 shows the in-plane electromechanical strain as a function of applied external electric field. It appears to be a parabolic relation with an obvious shift toward the positive electric field strength side. We believe that the observed strain should originate from a combination of electrostriction and piezoelectric effects. Fitting the non-linear strain-electric field relationship using a second order polynomial yields $\varepsilon_{11} = -0.00303E_3^2 + 0.0009E_3 - 1.79\times10^{-5}$ where $\varepsilon_{11}$ is the in-plane strain and $E_3$ is the strength of electric field along the perpendicular direction. The linear term arises from the piezoelectric effect. The deducted coefficient $d_{31}$ = 0.09 pm/V is shown in Fig. 2(b). It is consistent with the results of other GO molecules. It is well known that the electrostriction effect[38] has $\varepsilon_{11} = M_{13}E_3^2$ Thus the electric field-related electrostriction coefficient $M_{13} = -3\times10^{23}$ m$^2$/V$^2$ was obtained for our GO molecule. It is a surprise that a negative $M_{13}$ coefficient is obtained for our GO molecule because most electrostrictive polymers have a positive $M_{13}$. In other words, our GO molecule $C_{16}$O-sym-unzipped shows a contraction in the transverse direction upon the application of a perpendicular electrical field, whereas most electrostrictive polymers show a transverse elongation. Physical origins for such a distinctive electrostriction effect are not clear, which is worth of future investigation. Figure 4 indicates that the electrostriction dominate the electromechanical properties of sym-unzipped GO with a low oxygen concentration.

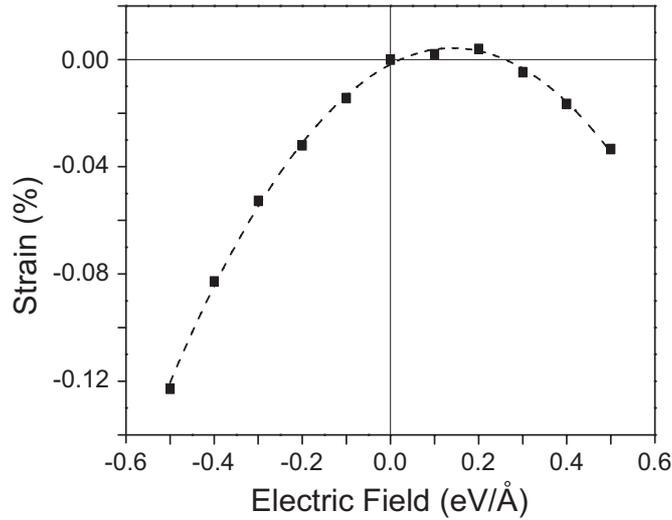

FIG. 4. In-plane strain of $C_{16}$O-sym-unzipped GO as a function of applied electric field perpendicular to the basal plane. A second order polynomial function (black dashed line) is fitted to the data obtained.

In summary, piezoelectric properties of GO with different structural configurations and oxygen concentration were studied using the first-principles density functional calculations. The maximal values of in-plane strain and strain piezoelectric coefficient were obtained for $C_2$O-sym-clamped GO, *i.e.*, a strain of up to 0.12% and $d_{31}$ = 0.24 pm/V. An increase of oxygen concentration in clamped GO enhances the piezoelectric strain output and $d_{31}$ coefficient, whereas an opposite trend is observed for unzipped GO. Through an in-depth structural analysis, we found that deformation of the oxygen-doped region dominates the piezoelectric strain of the sym-clamped GO. On the contrary, deformation from the 'graphene' region without oxygen dopant makes the major contribution to the piezoelectric strain of the unzipped GO. Interestingly, at a low oxygen concentration, the GO exhibits a much more profound electrostrictive deformation than the piezoelectric deformation. A negative electrostriction coefficient, $M_{13}$ is obtained by $C_{16}$O-sym-unzipped GO, in contrast with those of most electrostrictive polymers. Because of their excellent piezoelectric properties, robust molecular structures, and atomic thickness, GO molecules are promising two-dimension piezoelectric materials for MEMS/NEMS actuators and sensors.

**References**

1. Jeon, J.H., Cheng, T.H. & Oh, I.K. Snap-through dynamics of buckled IPMC actuator. *Sensors and Actuators A: Physical* **158**, 300–305 (2010).
2. Yeom, S.W. & Oh, I.K. A biomimetic jellyfish robot based on ionic polymer metal composite actuators. *Smart Mater. Struct.* **18**, 085002 (2009).
3. Rogers, G. W. & Liu, J. Z. Monolayer graphene oxide as a building block for artificial muscles. *Appl. Phys. Lett.* **102**, 021903 (2013).
4. Lu, J., Kim, S.G., Lee, S. & Oh, I.K. A Biomimetic Actuator Based on an Ionic Networking Membrane of Poly(styrene- alt-maleimide)-Incorporated Poly(vinylidene fluoride). *Adv. Funct. Mater.* **18**, 1290–1298 (2008).
5. Kim, U. *et al.* A transparent and stretchable graphene-based actuator for tactile display. *Nanotechnology* **24**, 145501 (2013).
6. Zhu, S.E. *et al.* Graphene-Based Bimorph Microactuators. *Nano Lett.* **11**, 977–981 (2011).
7. Mayer, A. C. *et al.* Polymer-based solar cells. *Materials Today* **10**, 28-33 (2007).
8. Baughman, R. H. Conducting polymer artificial muscles. *Synthetic Metals* **78**, 339–353 (1996).
9. de Lima, C. R. *et al.* A biomimetic piezoelectric pump: Computational and experimental characterization. *Sensors and Actuators A: Physical* **152**, 110–118 (2009).
10. Tabib-Azar, M. *Microactuators* (Kluwer Academic Publishers, Netherlands, 1997).
11. Novoselov, K. S. *et al.* Electric Field Effect in Atomically Thin Carbon Films. *Science* **306**, 666–669 (2004).
12. Geim, A. K. Graphene: Status and Prospects. *Science* **324**, 1530–1534 (2009).
13. Avouris, P., Chen, Z. & Perebeinos, V. Carbon-based electronics. *Nature Nanotech* **2**, 605–615 (2007).
14. Lee, C., Wei, X., Kysar, J. W. & Hone, J. Measurement of the Elastic Properties and Intrinsic Strength of Monolayer Graphene. *Science* **321**, 385–388 (2008).
15. Booth, T. J. *et al.* Macroscopic Graphene Membranes and Their Extraordinary Stiffness. *Nano Lett.* **8**, 2442–2446 (2008).
16. Pandey, D., Reifenberger, R. & Piner, R. Scanning probe microscopy study of exfoliated oxidized graphene sheets. *Surface Science* **602**, 1607–1613 (2008).
17. Yan, J.A., Xian, L. & Chou, M. Structural and Electronic Properties of Oxidized Graphene. *Phys. Rev. Lett.* **103**, (2009).
18. Baughman, R. H. *et al.* Carbon Nanotube Actuators. *Science* **284**, 1340–1344 (1999).
19. Bunch, J. S. *et al.* Electromechanical Resonators from Graphene Sheets. *Science* **315**, 490–493 (2007).
20. Xie, X. *et al.* An Asymmetrically Surface-Modified Graphene Film Electrochemical Actuator. *ACS Nano* **4**, 6050–6054 (2010).
21. Ong, M. T. & Reed, E. J. Engineered Piezoelectricity in Graphene. *ACS Nano* **6**, 1387–1394 (2012).
22. Prasad, R.K. *Quantum Chemistry* (New Age Science, UK, 2010).
23. Viefhues, M. *et al.* Physisorbed surface coatings for poly(dimethylsiloxane) and quartz microfluidic devices. *Anal Bioanal Chem* **401**, 2113–2122 (2011).
24. Tung, V. C., Allen, M. J., Yang, Y. & Kaner, R. B. High-throughput solution processing of large-scale graphene. *Nature Nanotech* **4**, 25–29 (2008).
25. Eda, G., Fanchini, G. & Chhowalla, M. Large-area ultrathin films of reduced graphene oxide as a transparent and flexible electronic material. *Nature Nanotech* **3**, 270–274 (2008).

26. Oh, J., Kozlov, M. E., Carretero-González, J., Castillo-Martínez, E. & Baughman, R. H. Thermal actuation of graphene oxide nanoribbon mats. *Chemical Physics Letters* **505**, 31–36 (2011).
27. Boukhvalov, D. W. & Katsnelson, M. I. Modeling of Graphite Oxide. *J. Am. Chem. Soc.* **130**, 10697–10701 (2008).
28. Rogers, G. W. & Liu, J. Z. Graphene Actuators: Quantum-Mechanical and Electrostatic Double-Layer Effects. *J. Am. Chem. Soc.* **133**, 10858–10863 (2011).
29. Rogers, G. W. & Liu, J. Z. High-Performance Graphene Oxide Electromechanical Actuators. *J. Am. Chem. Soc.* **134**, 1250–1255 (2012).
30. Xu, Z. & Xue, K. Engineering graphene by oxidation: a first-principles study. *Nanotechnology* **21**, 045704 (2009).
31. Kresse, G. & Furthmüller, J. Efficient iterative schemes for *ab initio* total-energy calculations using a plane-wave basis set. *Physical Review B* **54**, 169-185 (1996).
32. Kresse, G. & Joubert, D. From ultrasoft pseudopotentials to the projector augmented-wave method. *Physical Review B* **59**, 1758-1775 (1999).
33. Sun, G., Kertesz, M., Kürti, J. & Baughman, R. Dimensional change as a function of charge injection in graphite intercalation compounds: A density functional theory study. *Phys. Rev. B* **68**, (2003).
34. Zhang, Y. *et al.* Direct observation of a widely tunable bandgap in bilayer graphene. *Nature* **459**, 820–823 (2009).
35. Shimada, K. First-Principles Determination of Piezoelectric Stress and Strain Constants of Wurtzite III-V Nitrides. *Jpn. J. Appl. Phys.* **45**, L358–L360 (2006).
36. Hangleiter, A., Hitzel, F., Lahmann, S. & Rossow, U. Composition dependence of polarization fields in GaInN/GaN quantum wells. *Appl. Phys. Lett.* **83**, 1169 (2003).
37. Dekkers, M. *et al.* The significance of the piezoelectric coefficient $d_{31,eff}$ determined from cantilever structures. *J. Micromech. Microeng.* **23**, 025008 (2012).
38. DORF, R.C. *The Electrical Engineering Handbook* (CRC Press Inc, US, 1997).